\documentclass[%
 reprint,
superscriptaddress,
showpacs,
 amsmath,amssymb,
 aps,
 pre,
floatfix,
]{revtex4-1}

\usepackage[colorinlistoftodos]{todonotes}

\usepackage{graphicx}
\usepackage{subcaption}
\usepackage{dcolumn}
\usepackage{bm}

\begin{document}

\preprint{APS/123-QED}

\title{Granular front formation in free-surface flow of concentrated suspensions}

\author{Alessandro Leonardi}
\email{aleonardi@ethz.ch}
\email{Co-first author}
\affiliation{Institute for Building Materials, ETH Zurich\\
 Computational Physics for Engineering Materials\\
 Stefano-Franscini-Platz~3, 8093 Zurich, Switzerland
}
\affiliation{
Itasca Consulting GmbH \\
Leithestr. 111, 45886 Gelsenkirchen, Germany
}%


\author{Miguel Cabrera}
\email{miguel.cabrera@boku.ac.at}
\email{Co-first author}
\affiliation{
 Institute of Geotechnical Engineering\\
 University of Natural Resources and Life Sciences~(BOKU)\\
 Feistmantelstr.~4, 1180 Vienna, Austria
}%

\author{Falk K. Wittel}
\affiliation{Institute for Building Materials, ETH Zurich\\
 Computational Physics for Engineering Materials\\
 Stefano-Franscini-Platz~3, 8093 Zurich, Switzerland
}%

\author{Roland Kaitna}
\affiliation{%
 Institute of Mountain Risk Engineering\\
 University of Natural Resources and Life Science~(BOKU)\\
 Peter Jordanstr.~82, 1190 Vienna, Austra
}%

\author{Miller Mendoza}
\affiliation{Institute for Building Materials, ETH Zurich\\
 Computational Physics for Engineering Materials\\
 Stefano-Franscini-Platz~3, 8093 Zurich, Switzerland
}%

\author{Wei Wu}
\affiliation{
 Institute of Geotechnical Engineering\\
 University of Natural Resources and Life Sciences~(BOKU)\\
 Feistmantelstr.~4, 1180 Vienna, Austria
}%

\author{Hans J. Herrmann}
\affiliation{Institute for Building Materials, ETH Zurich\\
 Computational Physics for Engineering Materials\\
 Stefano-Franscini-Platz~3, 8093 Zurich, Switzerland
}%

\date{\today}

\begin{abstract}
Granular fronts are a common yet unexplained phenomenon emerging during the gravity driven free-surface flow of concentrated suspensions. 
They are usually believed to be the result of fluid convection in combination with particle size segregation. 
However, suspensions composed of uniformly sized particles also develop a granular front.
Within a large rotating drum, a stationary recirculating avalanche is generated. 
The flowing material is a mixture of a visco-plastic fluid obtained from a kaolin-water dispersion, with spherical ceramic particles denser than the fluid. 
The goal is to mimic the composition of many common granular-fluid materials, like fresh concrete or debris flow. 
In these materials, granular and fluid phases have the natural tendency to segregate due to particle settling. 
However, through the shearing caused by the rotation of the drum, a reorganization of the phases is induced, leading to the formation of a granular front. 
By tuning the material properties and the drum velocity, it is possible to control this phenomenon.
The setting is reproduced in a numerical environment, where the fluid is solved by a Lattice-Boltzmann Method, and the particles are explicitly represented using the Discrete Element Method. 
The simulations confirm the findings of the experiments, and provide insight into the internal mechanisms.
Comparing the time-scale of particle settling with the one of particle recirculation, a non-dimensional number is defined, and is found to be effective in predicting the formation of a granular front.

\end{abstract}

\maketitle

\section{Introduction}
 
Concentrated suspensions are a prominent multiscale material, since they exhibit a wide range of physical phenomena, whose relative importance varies widely at different length scales. 
The study of concentrated particle-fluid suspensions~\cite{Coussot1999,Ancey_2007}, is an essential research field due to its many applications, among which the processing of food and pharmaceutical materials, the construction industry, and the study of natural hazards are notable examples.
Among those, phase separation is typical of suspensions where the length scale is sufficiently large so that particles are able to develop inertia independently of the fluid. This has dramatic and potentially undesirable effects, since the material loses its uniformity and becomes therefore problematic to process. The modeling of phase separation is still a challenge, both numerically and analytically, since constitutive relationships able to describe the resulting non-homogeneous material have difficulties achieving generality. 

Particle suspensions with grains heavier than the fluid vertically segregate due to sedimentation. The same suspension, if flowing down an incline, can also experience other segregation patterns. Examples are inverse and lateral grading as a function of the confinement conditions (2D or 3D flow).
In many cases, a concentration of particles at the front of the flow is observed. 
This behavior is well documented~\cite{Gray2009,Johnson2012a,Marks2011} and can be observed in nature~\cite{Pierson1986}. A particularly important example are debris flows, an avalanche that originates in mountainous terrain composed of a mixture of water with granular material~\cite{Hungr_2001}. 
For still debated reasons, the front of the debris flows is often richer in large boulders than the rest of the flow. This granular front brings a significant contribution to the destructive potential a debris flow represents.
Similar mechanisms are also observed in snow avalanches~\cite{Nohguchi2009}, pyroclastic flows, and rock slides. At smaller scales, grain-rich fronts have been observed on thin film particle-laden flow~\cite{Zhou2005,Aranson2006}.
The formation of a granular front is usually explained as an effect of inverse size segregation~\cite{Johnson2012a,Gray2011}: When a mixture of particles flows down an incline, the shear deformations causes an increase in the porosity of the medium, and a number of void spaces are continuously created. If the particles that populate the system are not of the same size, it is more likely that smaller particles will fall into the newly created voids. As a consequence, larger particles tend to migrate towards the surface where they are advected by the flow and brought towards the front. 
However, the presence of particles of different sizes is not necessary to generate a particle-rich front. Kaitna et al.~\cite{Kaitna2011} showed that the same pattern can be observed with particles of uniform size immersed in a fluid. They studied how the flow characteristics influence the formation of a granular front, showing that the phenomenon is dependent on the particle content, fluid viscosity, and on the shear conditions. This scenario however is much less studied due to the difficulties in performing experiments and simulations. Thus the mechanism that induces the formation of granular fronts remains unclear. 

\begin{figure}[t]
 \centering
 \includegraphics[scale=1]{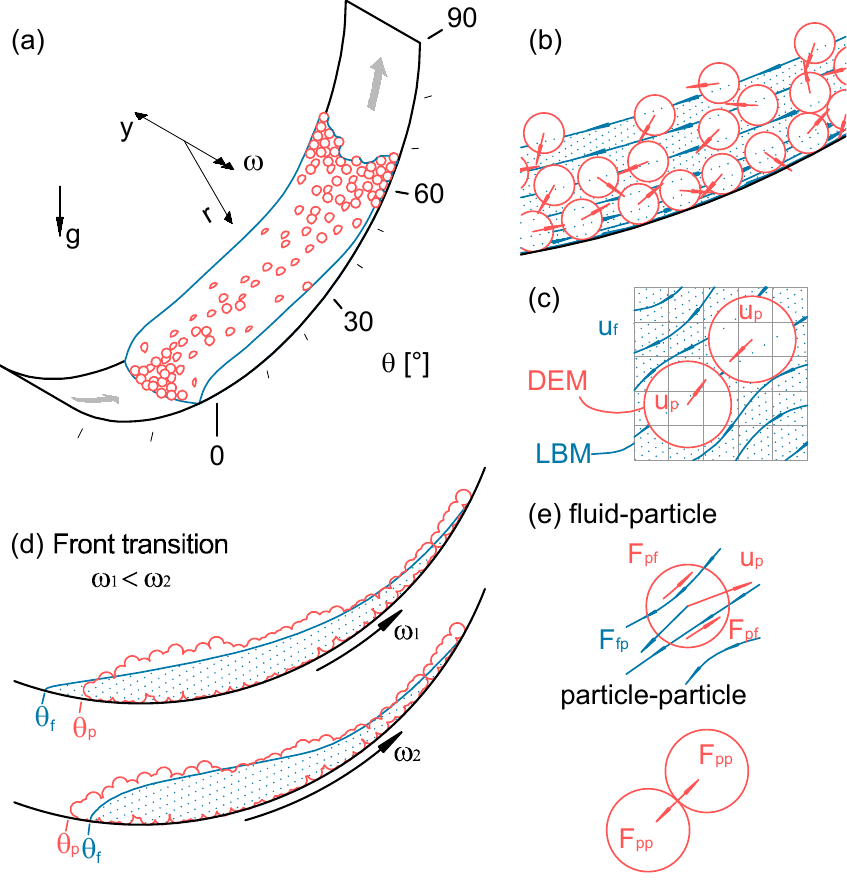}
 \caption{Schematics of the particle fluid mechanisms occurring in a rotating drum test. The color scheme is blue for the fluid and red for the particles. (a) Rotating drum reference system and common particle-fluid distribution. (b) Particle and fluid motion patterns. (c) LBM and DEM discretization mesh employed in the numerical simulations (a two-dimensional view is presented for simplicity). (d) Granular front formation process by increasing the rotational velocity of the drum $\omega_i$. (e) Fluid-particle and particle-particle interactions considered as a result of a granular flow through a viscous flowing fluid.}
 \label{fig:graphicalAbstract}
\end{figure}

With the use of a rotating drum~\cite{Hill2001,Pignatel2012,Chou2012a,Turnbull2011,Liao2014,Yang2003,Jain2002,Ding2001,Search1995,Baumann1995,Search1994} (see Figs.~\ref{fig:graphicalAbstract}~(a),~\ref{fig:Rot_drum}), stationary, free-surface recirculating flows are obtained, whose physical properties can be continuously measured. The segregation pattern can be induced and controlled by regulating the rotational velocity of the drum as well as by modifying the particle concentration of the mixture (see Fig.~\ref{fig:graphicalAbstract}~(d)). 
In order to gain insight into the mechanisms that govern the transition to a granular front, the experimental settings are reproduced in a numerical environment. The used method is based on the well-established coupling between the Discrete Element Method (DEM) for the representation of the particle phase, and the Lattice-Boltzmann Method (LBM) for the solution of the fluid phase. 
 
This paper offers a description of the experimental (Sec.~\ref{sec:experiments}) and numerical (Sec.~\ref{sec:numerical}) setting, followed by the presentation of the results in Sec.~\ref{sec:results} and their interpretation in Sec.~\ref{sec:analytical}. Sec.~\ref{sec:conclusions} summarizes and comments on the conclusions.


\section{Experimental methods}
\label{sec:experiments}

\begin{figure}[t]
 \centering
\includegraphics[]{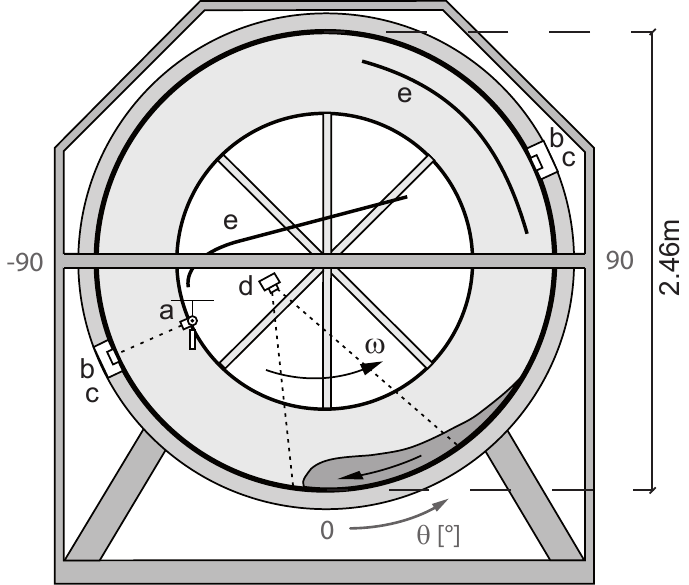}
 \caption{Rotating drum at BOKU. 
a)~Height point-laser, b)~Normal load cell, c)~Fluid pressure transducer, d)~High-speed camera, and e)~Inner-roofs (drop protection). Adapted from Refs. \cite{Kaitna_2007a,Schneider_2011}.}
 \label{fig:Rot_drum}
\end{figure}

The rotating drum employed throughout this study is located at the University of Natural Resources and Life Sciences (BOKU), Vienna, Austria, and described in detail in Refs.~\cite{Kaitna_2007a,Kaitna_Rickenmann_2007,Cabrera_etal_2014b}. Therefore, only a brief description of the instrumentation and the configuration of the drum for the experiments in this study is presented. The rotating drum has a diameter of $R=2.46~\text{m}$ and a rectangular inner section of width $W=0.45~\text{m}$ (see Fig.~\ref{fig:Rot_drum}). 
The cylindrical reference system used in this paper is presented in Fig.~\ref{fig:Rot_drum}, with the angular coordinate $\theta$ being 0$^\circ$ at the 6 o'clock position and 90$^\circ$ and -90$^\circ$ at the 3 and 9 o'clock position, respectively. 
The side-walls are composed at one side of stainless steel and at the opposite side of acrylic glass, while the bed of the channel is roughened using a rubber surface with protrusions in a zigzag pattern of approximately 3~mm in height and 5~mm in separation. 
Material losses are observed due to the clogging of material in the roughened bed, which then drops and accumulates at the inner roofs (Fig.~\ref{fig:Rot_drum}e). The loss of material is recognized as one of the major limitations of the tests conducted, and is noticeable for high viscous flows. The drum rotates around its axis at a constant rotational speed $\omega$ of approximately $0.3$, $0.5$, $0.7$, or $0.9~\text{rad}/\text{s}$.
%
%
%

The material is chosen to represent a simplified natural suspension, resembling to some extent the one in a debris flow. Particles are ceramic balls of density $\rho_\text{s}= 2420~\text{kg}/\text{m}^3$ and diameter $d=~32.6 \pm0.03~\text{mm}$, whose high strength and stiffness enables the occurrence of strong collisional forces without large deformations or material failure.
The friction angle between the ceramic balls and the drum roughened bed material is $\psi_\text{bed}=42.5^\circ$, measured in a similar way as presented in Ref.~\cite{Henein_1980}, while the inter-particle friction angle is $\psi_\text{s}=27.7^\circ$~\cite{Wu2010}. 
Particles collide with a constant coefficient of restitution of $c=0.7$ when dry, and $c\simeq0$ when covered with fluid. 

\begin{figure}[t]
\centering
\includegraphics[scale=1]{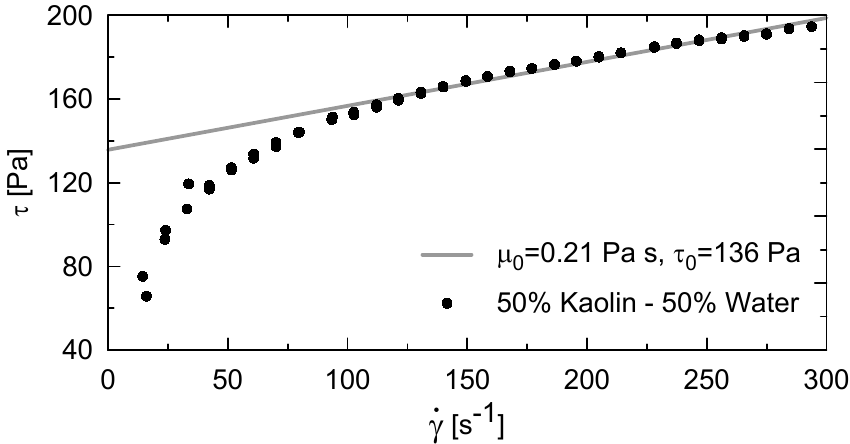}
\caption{Rheological behavior of the kaolin-water dispersion. The flow curve is approximated as a Bingham fluid (solid line).}
\label{fig:Kaolin_rheology}
\end{figure}

The fluid phase is obtained by mixing kaolin powder (particle size of $2.0~\mu\text{m}$~($D_{50}$), and particle density of $2600~\text{kg}/\text{m}^3$) with water. Changing the relative proportions of kaolin and water, mixtures of different rheological behaviors are obtained~\cite{Coussot1997}. The parameters of the kaolin-water dispersion are measured using a simple co-axial cylinder rheometer (Bohlin Visco~88) with a gap of $1.5~\text{mm}$. 
The mixture used in the experiments is composed of kaolin and water in equal parts by mass, resulting in a fluid with density $\rho_\text{f}=1420~\text{kg}/\text{m}^3$ and a rheological flow curve as presented in Fig.~\ref{fig:Kaolin_rheology}. For simplicity in the  numerical analysis (Sec.~\ref{sec:numerical}, \ref{sec:results} ), the resulting dispersion is assumed to behave like a plastic fluid with a yield stress $\tau_0$, whose constitutive relation between shear stress $\tau$ and shear rate $\dot{\gamma}$ is approximated with a Bingham law
\begin{equation}
\tau=\tau_0 + \mu_0 \dot{\gamma},
\end{equation}
where $\mu_0$ is the plastic viscosity (see Fig.~\ref{fig:Kaolin_rheology}). 

During the experiments, a volume (net of losses) of $40~\text{kg}$ of fluid is added to the drum, together with a variable amount of particles ($10$, $20$, $30$, $40$, $50~\text{kg}$). This results in mixtures with a global particle concentration in the range $13$, $23$, $31$, $38$ and $43\%$  respectively. A couple of initial rotations are completed to uniformly cover the channel bed and reach steady state.

The drum is instrumented as shown in Fig.~\ref{fig:Rot_drum}. At every successive rotation, basal total load and basal fluid pressure are measured with a set of two load cells (HBM PW2GC3) and one piezoresistive pressure transmitter (Keller PR25Y), fixed normal to the bed and displaced 180$^\circ$ along the circumference. The load cells are connected to a plate, $60~\text{mm}$ in diameter, covered by the same roughened layer of the drum's bed. At the same time, the flow height is recorded by a point-laser sensor (Baumer OADM~20), rotating with the drum yielding a complete height profile every turn of the drum. All instruments record at a sampling frequency of 1200~Hz. In addition, a high-speed camera (Optronis CR3000x2) with a $28$-$200~\text{mm}$ lens records the front of the flowing material. The camera is mounted near the center of rotation of the drum, with a focal distance of $1.2~\text{m}$ to the channel bed. The videos are recorded at a frame rate of $500~\text{fps}$, with a frame size of $1696\times1710~\text{pixels}$. The images allow to reconstruct the front location and to identify the position of the particles (see Sec.~\ref{sec:results}).

The material inside the drum forms a stationary free-surface flow, whose front, body, and tail present distinct features. Along the channel axis, the front presents velocities both in the $\theta$ and $r$ directions, while in the body, velocities in $r$ are negligible. At the tail the flow gives rise to unsteady avalanche releases~\cite{Henein_etal_1983a}.


\section{Outline of the LBM-DEM method}
\label{sec:numerical}
\begin{figure}[t]
\centering
\includegraphics[scale=1]{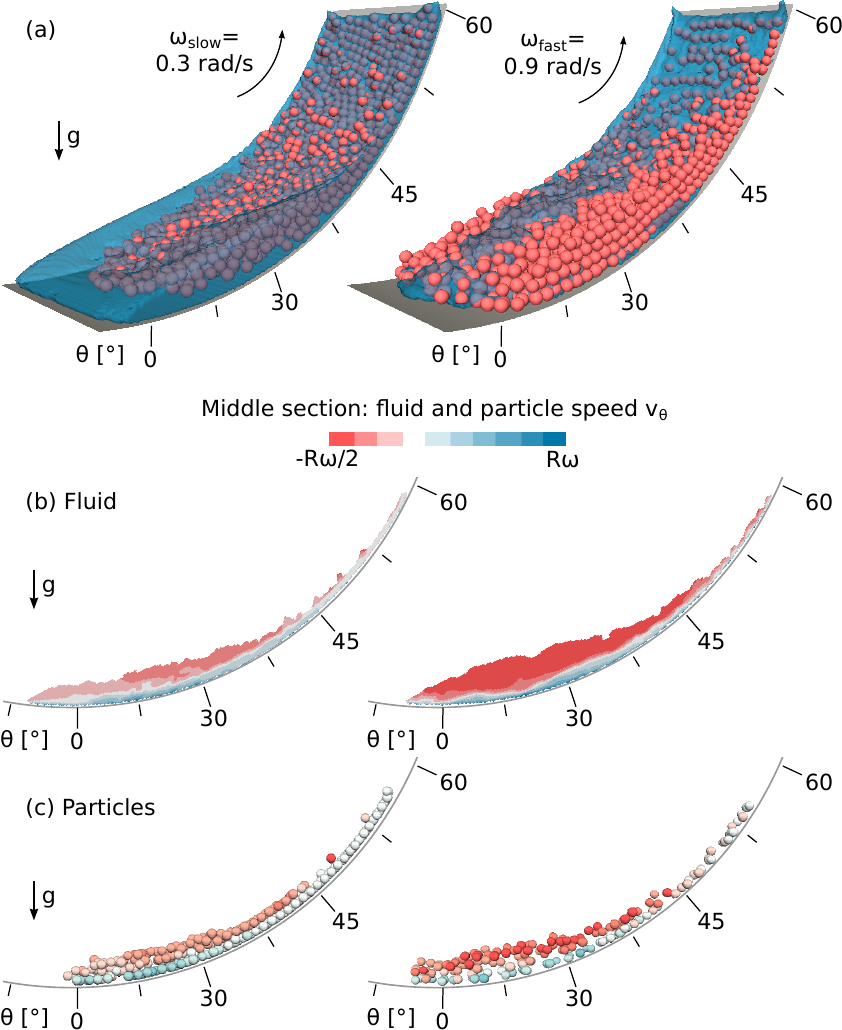}
\caption{Snapshots of the DEM-LBM simulations (particle content of $31\%$). (a) The formation of a granular front by an increase in the angular velocity of the drum. The tangential velocity $v_\theta$ along a longitudinal profile for fluid (b) and particles (c) . }
\label{fig:simulations}
\end{figure}
The experimental setting described in the precedent chapter is recreated in a numerical environment. The natural choice for the simulation of the particle-fluid mixture is a hybrid DEM-LBM method, due to its flexibility to handle the many challenges posed by the non-linear boundary conditions. 
A detailed presentation of the DEM-LBM scheme is beyond the scope of the present work. 
The readers are referred to the literature on the LBM 
in Refs.~\cite{Chen1998,Succi2001} and on the DEM in Ref.~\cite{Bicanic2007}. The coupling between these two solvers has been the object of much recent research~\cite{Aidun2010} and has nowadays achieved a high level of maturity, with the possibility to simulate real systems also with non-Newtonian fluids~\cite{Leonardi2014}. 

The domain is discretized in a fixed, regular, cubic lattice, whose nodes belong to these five categories: fluid, gas, interface, solid walls, or solid particle. The role and governing equations for every node type are described below.

\textbf{Fluid nodes} are the locations where the fluid dynamics is solved. This is realized through a discretization of the Boltzmann equation. The fluid is not represented by continuous fields of macroscopic variables, like in most classic solvers, but rather through streams of microscopic particles, or populations, whose collisions and subsequent redistribution governs the dynamics.
 The population density $f$ at every fluid node is used to reconstruct density $\rho_\text{f}$, pressure $p_\text{f}$, and velocity $\boldsymbol{u}_\text{f}$ through simple summations. If $f_i$ describes a population moving at speed $\boldsymbol{c}_i$, this translates into
 \begin{eqnarray}
\label{densityVelocity}
 \rho_\text{f}=\sum\limits_i f_i, \ &
 p_\text{f}=c_s^2 \cdot \rho_\text{f}, \ &
 \boldsymbol{u}_\text{f}=\sum\limits_i f_i\boldsymbol{c}_i/\rho_\text{f},
\end{eqnarray}
which implies that the fluid is actually treated like a slightly compressible medium, and the pressure is a secondary variable obtained by multiplication of $\rho_\text{f}$ with the square of the system speed of sound $c_s^2$.
The evolution of $f_i$ during a unitary time step is governed by the Lattice-Boltzmann equation
\begin{eqnarray}
\label{newfunctions}
f_i (\boldsymbol{x}+ \boldsymbol{c}_i,t+1) =& \nonumber \\
f_i (\boldsymbol{x},t)&+\Omega_i (\boldsymbol{x},t)+ F_i (\boldsymbol{x},t,\boldsymbol{g}+\boldsymbol{p}),
\end{eqnarray}
where $F_i$ implements the effect of gravity $\boldsymbol{g}$ and of the fluid-particle coupling term $\boldsymbol{p}$, while $\Omega_i$ is an operator describing the effect of population collisions. Commonly expressed by the Bhatnagar-Gross-Krook linear approximation~\cite{Bhatnagar1954}, it relaxes the system to an equilibrium state $f^{eq}_i$,
\begin{eqnarray} \label{collision}
\Omega_i=\frac{f_i^{\text{eq}}-f_i}{\tau}.
\end{eqnarray}

The transition to equilibrium is described by the relaxation time $\tau$, which is proportional to the plastic viscosity of the fluid $\mu_0$  and its yield stress $\tau_0$~\cite{Leonardi2014a} as
\begin{eqnarray}
\label{mu}
\tau=\frac{1}{2}+\frac{\tau_0/\dot{\gamma}+ \mu_0}{c_s^2}.
\end{eqnarray}

\textbf{Gas nodes} represent the space not occupied by the fluid, and therefore neither contain nor transfer populations.

\textbf{Interface nodes} represent the interface between fluid and gas, and are similar to fluid nodes, in the sense that the streaming of population happens in an identical fashion. However, they are granted an additional degree of freedom, a variable called mass $m \in (0,1)$, used to track the evolution of the surface. Interface nodes mutate into fluid nodes if $m\geq1$ or into gas nodes if $m\leq0$. The evolution of mass depends on the difference between the populations streaming in and out of the node,
\begin{eqnarray}
m(t+1) = m (t) +\sum \alpha \left( f_\text{in} - f_\text{out} \right)
\end{eqnarray}
where $\alpha$ depends on whether the population exchange happened between two interface nodes or between a fluid and an interface node~\cite{Korner2005}.
The modeling of surface tension can be included in this formulation. However for the studied cases a set of test simulations revealed surface tension to have no sensible effect on the results, and its modeling has therefore been neglected.

\begin{table}[b]
\caption{\label{tab:particleParameters} Material parameters used in the DEM.}
\begin{ruledtabular}
\begin{tabular}{l c}
     Parameter & Value\\ \hline
    Mass density & $2420\ \text{kg}/\text{m}^3$ \\  
    Young's modulus & $1.6\cdot 10^7\ \text{Pa}$ \\ 
    Poisson's ratio & $0.3$ \\  
    Restitution coefficient & $\simeq 0$  \ \\  
    Shear damping & $10\ \text{kg}/\text{s}$ \\  
    Particle-particle friction angle & $27.7^\circ$ \\  
    Particle-drum friction angle & $42.5^\circ$ \\  
\end{tabular}
\end{ruledtabular}
\end{table}

\textbf{Solid wall nodes} are all nodes that lie inside solid walls, i.e. the drum cylinder and lateral walls. They do not contain nor transfer populations, but nevertheless affect the fluid since they both enforce no-slip at the boundary and transfer momentum to the fluid when the drum is in motion. No-slip is enforced by requiring all population streams $f_{i}$ pointing towards a solid node to be reflected back. The reflected populations $f_{i'}$ are modified taking into account the momentum transfer,
\begin{eqnarray}
\label{SBB}
 f_{i'}= f_{i} - 6 w_i \rho_\text{f} \boldsymbol{u}_\text{w} \cdot \boldsymbol{c}_i,
\end{eqnarray}
where $\boldsymbol{u}_\text{w}$ is the velocity of the wall at the reflection location and $w_i$ is a weight depending on the lattice configuration.
Nodes belonging to the drum surface employ a modified version of this rule, to take into account the curvature of the boundary. Details about the curved-surface treatment can be found in Refs.~\cite{Mei2000,Mei2002}.

\textbf{Solid particle nodes} are all nodes contained inside solid particles. For the particle-fluid coupling algorithm the direct forcing approach~\cite{Feng2005} is employed in a simplified form similar to the one described in Ref.~\cite{Leonardi2014b}. A sketch of the interaction mechanism is given in Fig.~\ref{fig:graphicalAbstract}~(c). Particles are immersed in the fluid and are advected through the LBM regular grid. The difference between particle and fluid velocity is used to compute a hydrodynamic interaction force that is transmitted both to the fluid and to the particle equations of motion. The Lattice-Boltzmann equation is solved in this case with the additional forcing term $\boldsymbol{p}$ appearing in Eq.~\ref{newfunctions}, calculated as
\begin{equation}
 \boldsymbol{p}(\boldsymbol{x},t)={\rho_\text{f}(\boldsymbol{x},t)}\left[ \boldsymbol{u}_\text{f}(\boldsymbol{x},t) - \boldsymbol{u}_\text{p}(\boldsymbol{x},t)\right],
\end{equation}%
where $\boldsymbol{u}_\text{p}$ is the velocity of the particle at the node location \cite{Kang2011}. A force of opposite sign is applied to the particle.

Particle dynamics is solved using a DEM model (see Fig.~\ref{fig:graphicalAbstract}~(e)). The collisions between particle, and the consequent momentum exchange is modeled using a Hertzian law~\cite{Brilliantov1996,Leonardi2014b} with constant coefficient of restitution. The material parameters used in the simulations are recapped in Table~\ref{tab:particleParameters}.

\section{Comparison between numerical and experimental results}
\label{sec:results}

\begin{figure}[t]
 \centering
 \begin{subfigure}[h]{1\columnwidth}
\includegraphics[scale=1]{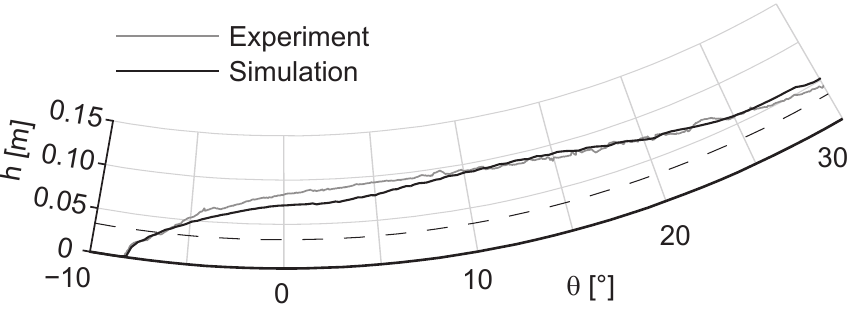}
\caption{}
 \label{fig:comparison_height}
\end{subfigure}
\centering
\begin{subfigure}[h]{1\columnwidth}
 \centering
\includegraphics[scale=1, angle=90]{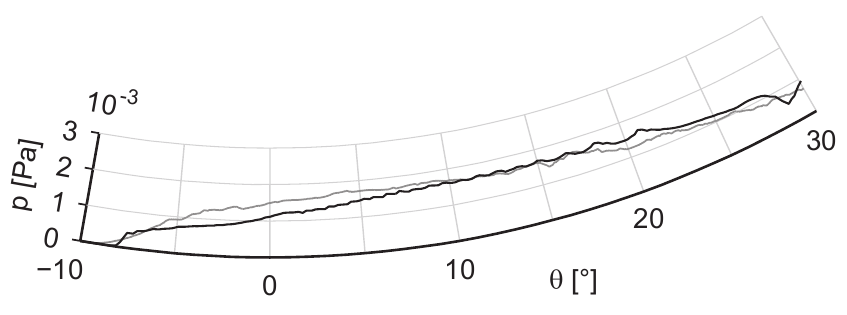}
\caption{}
 \label{fig:comparison_pressure}
\end{subfigure}
\caption{Experimental and numerical flow height (\ref{fig:comparison_height}) and bulk basal pressure profiles (\ref{fig:comparison_pressure}) at $0.3~\text{rad}/\text{s}$ and particle concentration of 31\%. The dashed line has the height of one particle diameter.}
\label{fig:exp_vs_sim_height_pressure}
\end{figure}

The  height and the basal pressure longitudinal profiles obtained in experiments described in Sec.~\ref{sec:experiments} are used to validate the DEM-LBM scheme. Fig.~\ref{fig:comparison_height} presents the height profile as detected by the point-laser sensor and its numerical counterpart. Both profiles are averaged over all available data in steady state. The experimental profile is based on $6$ recorded data sets, one for every full rotation of the drum, while the numerical profiles are obtained from a much larger set of data. Good agreement is found between the numerical simulation and the experimental measurements, matching reference points as the front location and flow height. In a similar way, the experimental basal pressure profile (Fig.~\ref{fig:comparison_pressure}) coincides with the numerical pressure, accounting simultaneously for the particle and fluid contributions.

\begin{figure}[t]
 \centering
\includegraphics[width=1\columnwidth]{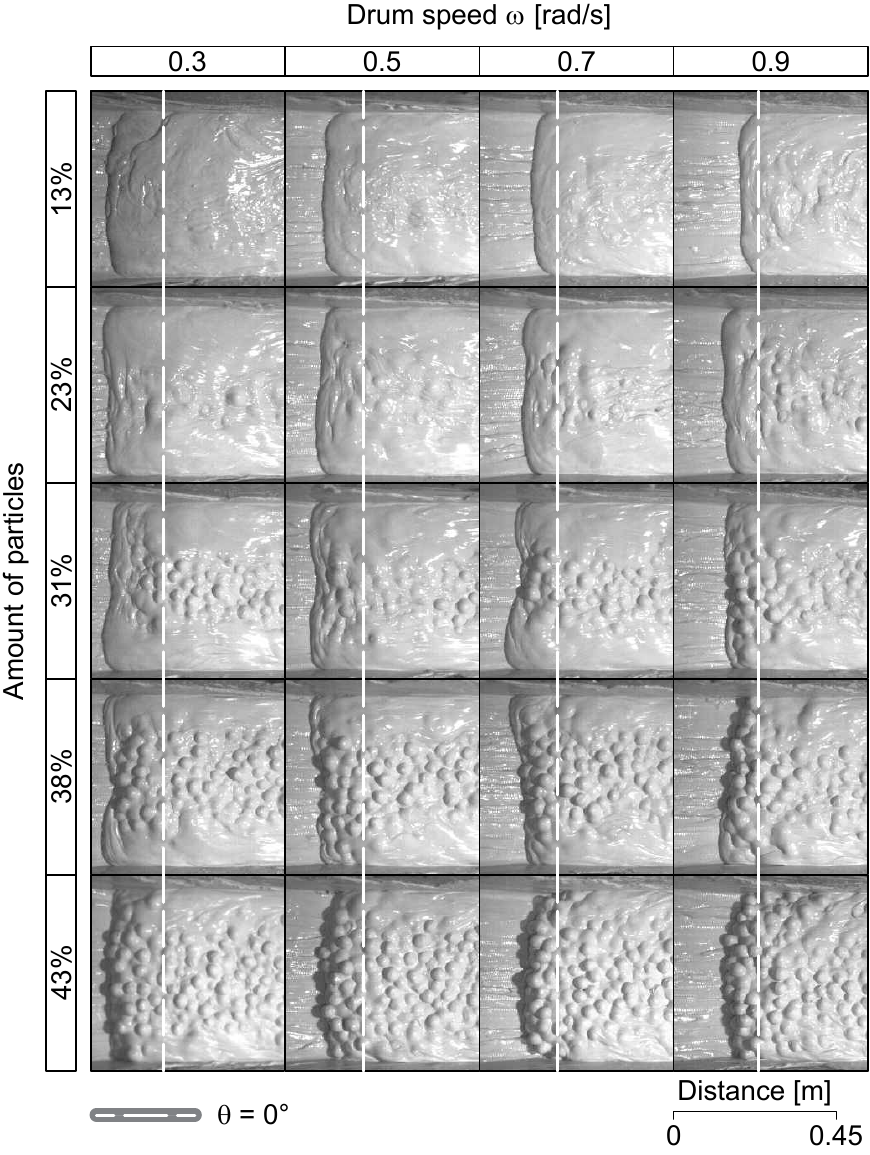}
 \caption{Front as a function of the angular velocity of the drum and particle concentration.}
 \label{fig:front_location_experimental}
\end{figure}
\begin{figure*}[t]
 \centering
\includegraphics[]{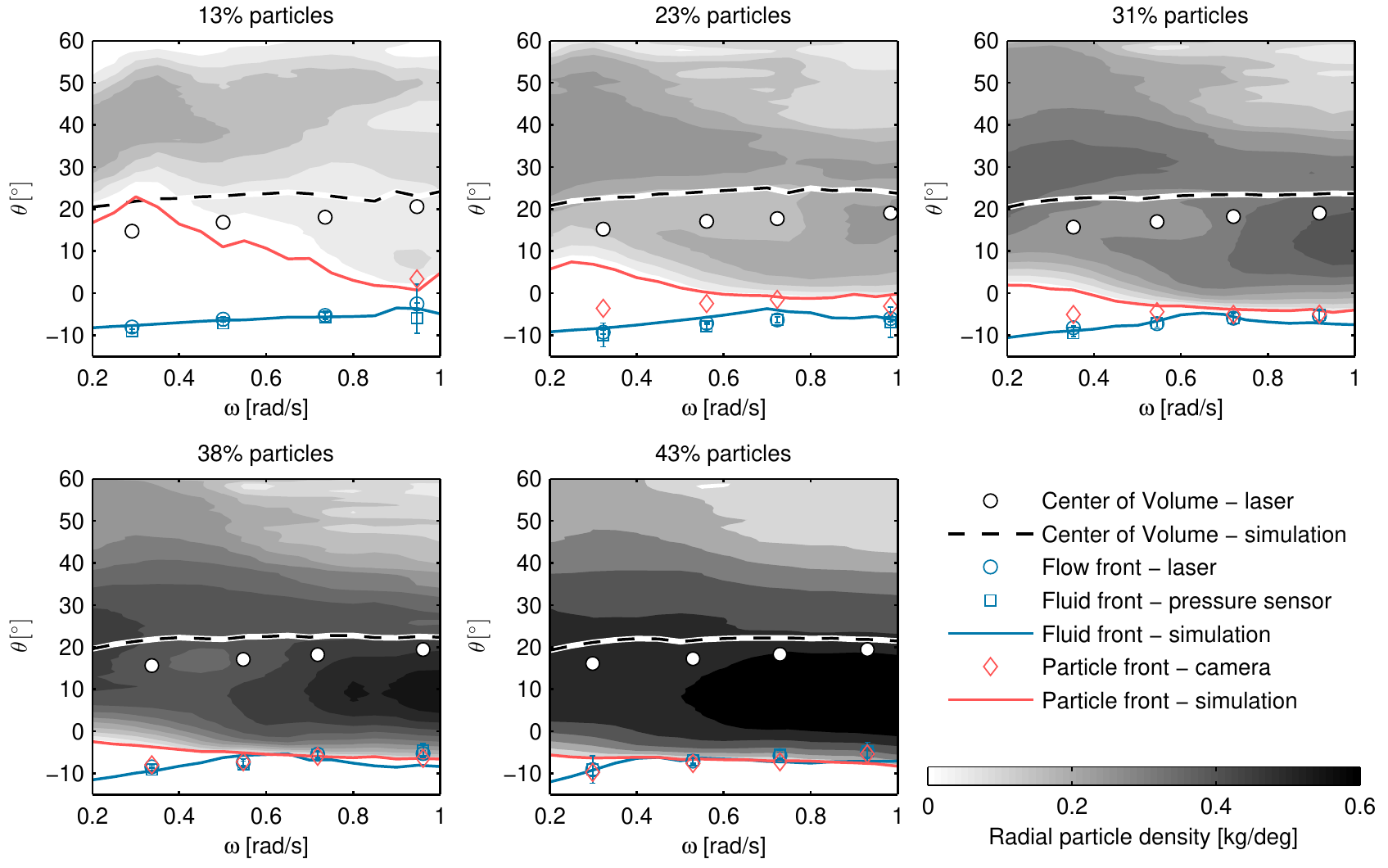}
 \caption{Phase front transition. Experimental and numerical simulation results are shown. The experimental fluid front is extracted from the recorded height profiles, while the experimental particle front comes from the analysis of the the images obtained with the high-speed camera.}
 \label{fig:averageFronts}
\end{figure*}

The position of the particles inside the fluid cannot be clearly obtained with the sensors. Therefore the sequence of frames obtained with the high-speed camera (see Fig.~\ref{fig:front_location_experimental}) is analyzed with an edge-detection technique~\cite{Canny_1986} for the flow front recognition. The most striking characteristic of the flow is the possibility to obtain a granular front, even with a relatively low initial concentration of the particles. The higher the rotational speed of the drum, the more particles will get close to the fluid front. The formation of a granular front strikingly resemble the one observed in a debris flow. The process is shown in Fig.~\ref{fig:averageFronts}, where the fluid front position $\theta_\text{f}$, as identified by the sensor is shown with blue markers, while the particle front position $\theta_\text{p}$ as obtained by the high-speed camera, is shown with red markers. Fig.~\ref{fig:averageFronts} also shows the position of the center of volume (CoV), obtained by integration of the height profiles. The values obtained from the simulations are shown with continuous lines.

In agreement with Ref.~\cite{Kaitna2011}, the transition from a fluid front into a particle front can be induced by an increase in the angular velocity of the drum (see Fig. \ref{fig:front_location_experimental}). At low angular velocities, the particles are concentrated at higher inclinations.
Then, with increasing velocities at the base, the particles diffuse, and start to move towards lower inclinations. 
Along with the motion of the particles, the interstitial fluid is strongly sheared by the particle motion and drum bed.
As a result of this shearing, the fluid front moves backwards while the particles move forward until the particle front is formed.

\section{Internal mechanisms governing phase separation}
\label{sec:analytical}

The emergence of granular fronts, described in the previous section, is the result of two mechanisms. The first is the migration of the fluid front towards higher positions, observed for growing angular velocity of the drum, and the second is the movement of the particles under high shear. Simple analytical models for these two phenomena are described in the following.

\begin{figure}[t]
 \centering
 \includegraphics[scale=1]{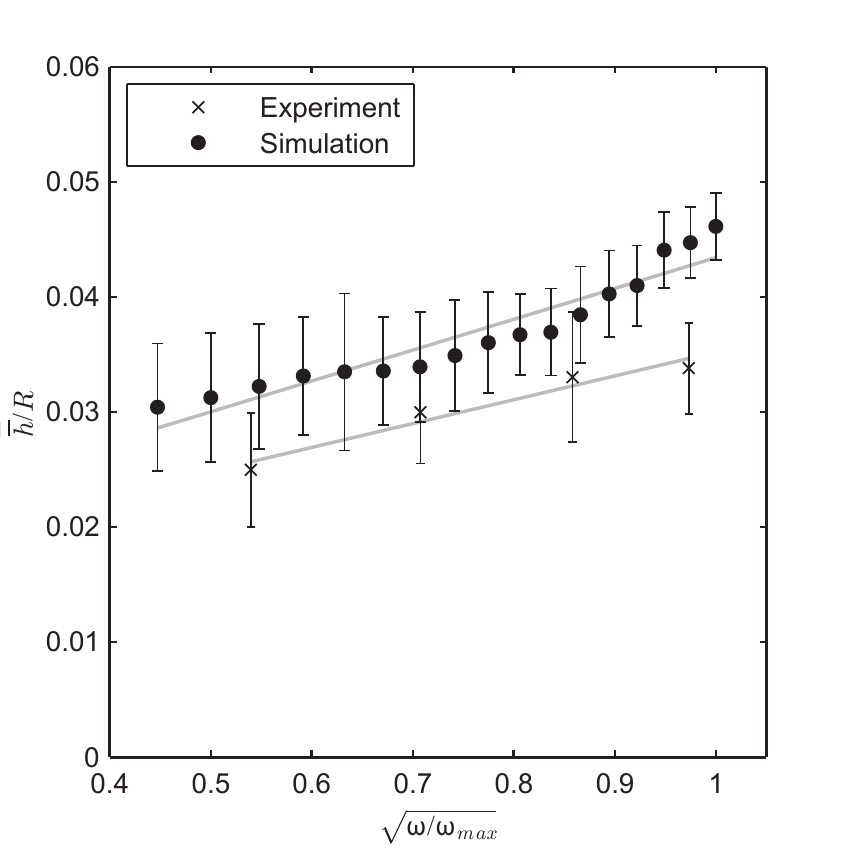}
 \caption{Analytical flow height of a free surface flow down an incline against the experimental and numerical height (average height between 25 to 30$^\circ$).}
 \label{fig:heightVsSpeed}
\end{figure}

As seen in Fig~\ref{fig:Kaolin_rheology}, the Kaolin-water mixture can be approximated with a Bingham fluid, characterized by a yield stress $\tau_0$ and a plastic viscosity $\mu_0$.  Obtaining an analytical solution of the Navier-Stokes equations for a Bingham plastic in the partially-filled rotating drum is not an easy task: The fluid presents a free surface whose shape is not given a priori.  However, considering a single slice of the drum ($\theta$ fixed) and approximating it with a one dimensional flow in the $\theta$ direction, an approximate solution can be obtained, which is analogous to the use of a long-wave approximation~\cite{Liu}. With this hypothesis the flow is assumed to be steady and self-similar over the width $w$ of the drum and the edge effects due to the side walls of the drum are neglected. The shear rate and shear stress tensors each boil down to one component $\dot{\gamma}_{\theta r} = \frac{\partial{v_\theta}}{\partial r}$ and  $\tau_{\theta r}$. 
The Bingham constitutive law can be written in this new system as:
\begin{equation}
 \left\{
  \begin{array}{l l}
    \left|\frac{\partial{v_\theta}}{\partial r}\right|= 0 & \quad \textrm{if}\ (|\tau_{\theta r}|<\tau_0),\\
    \tau_{\theta r} = \left( \frac{\tau_0}{ \left|\frac{\partial{v_\theta}}{\partial r}\right|} + \mu_0 \right) \frac{\partial{v_\theta}}{\partial r}  & \quad \textrm{if}\ (|\tau_{\theta r}|>\tau_0).
  \end{array}
  \right.\
\end{equation}
The integration of the momentum equation with limits $r=R$ at the base of the drum, and $r=r_h$ at the free surface position, gives the solution of the tangential-velocity profile  $v_\theta(r)$
\begin{equation}
  \begin{array}{l l}
    \omega R-a\sin \theta (r-R)(r+R-2r_0) & \quad \textrm{if } \ r_0<r<R,\\
    \omega R-a\sin \theta (R-r_0)^2 & \quad \textrm{if } \ r_h<r<r_0,\\
  \end{array}
  \label{u_prof}
\end{equation}
where $\omega$ is the rotational speed of the drum, and $a=\rho_\text{f} g / 2\mu_0$. Between the positions $r_0$ and $r_h$ a plug flow forms, whose total width $h_0=r_0-r_h$ is proportional to the yield stress as:

\begin{equation}
 h_0=\frac{\tau_0}{\rho_\text{f} g \sin \theta}.
\end{equation}
By imposing stationary conditions, and therefore total flux $q_\theta=0$ in the $\theta$ direction, the total height of the fluid $h=R-r_h$ is obtained. The following expression can be obtained by integrating the velocity profile (Eq.~\ref{u_prof}) over the whole height $h$ (both yielded and unyielded regions):
\begin{equation}
 q_\theta=\omega R h -\frac{1}{3} a\sin\theta\left(h-h_0\right)^2 \left(2 h+h_0\right)=0.
\end{equation}
This equation can be solved numerically to obtain the height $h$ of the flow. An elegant solution can be extracted with the approximation $2h+h_0 \approx  3h$, which is justified by $h_0$ being of the same order of $h$. Thus
\begin{equation}
h=h_0 +\left(\frac{\omega R }{a \sin{\theta}}\right)^{1/2}.
\label{eq:anProf}
\end{equation}
This solution does not hold for positions in the flow where a non-negligible $r$ component exists, like the flow front. In addition, it is a general solution that does not take into account the actual flowing volume. However, for positions in the drum sufficiently far from the front ($\theta>10^\circ$), Eq.~\ref{eq:anProf} correctly captures the trend of variation of the height as linear function of the square root of the drum rotational speed $\sqrt{\omega}$. This is so because at higher basal velocities, in order to keep a stationary motion, the fluid needs to accumulate mass by increasing its height, an effect typical for free-surface flows of shear-dependent fluids. Fig~\ref{fig:heightVsSpeed} shows the fluid height  measured in experiments and simulations as a function of $\sqrt{\omega/\omega_\text{max}}$ (with $\omega_\text{max}=1.0~\text{rad}/\text{s}$). The average height $\bar{h}$ in the range of $25^\circ \leq \theta \leq 30^\circ$ is plotted, showing a good agreement with the linear trend predicted by Eq.~\ref{eq:anProf}.

\begin{figure}[t]
 \centering
 \includegraphics[scale=1]{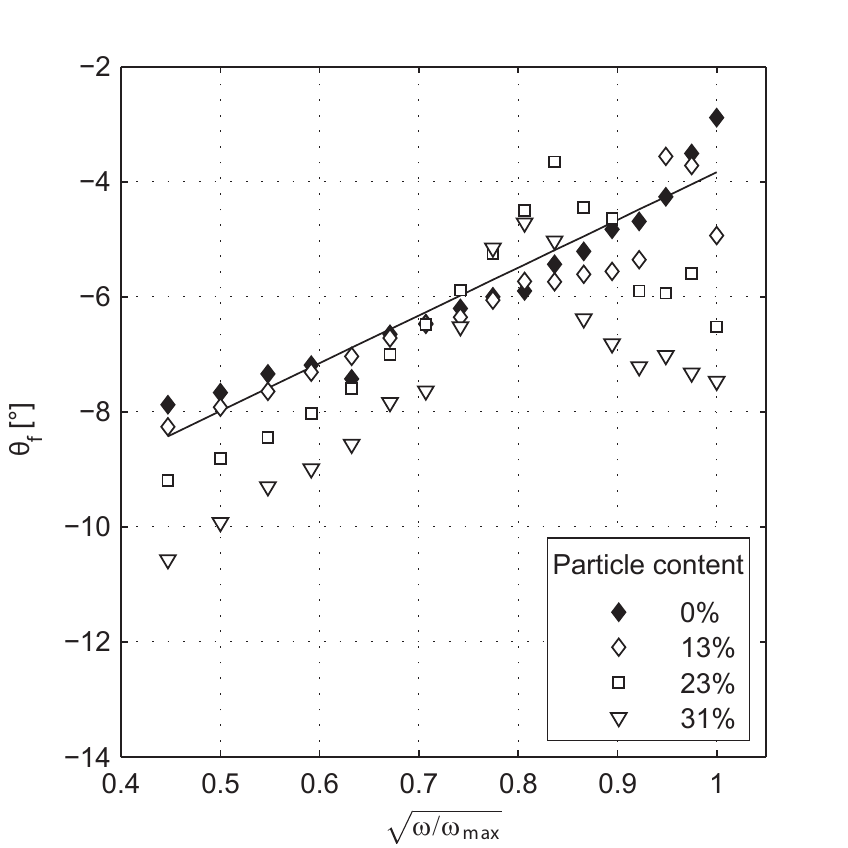}
 \caption{Numerical location of the fluid front $\theta_\text{f}$ as a function of the angular velocity of the drum $\omega$. Markers represent different particle contents. For a fluid-only simulation the trend is linear (solid line).
 }
 \label{fig:FluidFrontVsSpeed}
\end{figure}

Since the flowing volume is constant, the rise of the fluid height given by Eq.~\ref{eq:anProf} must be accompanied by a transfer of mass from the front. This explains why the fluid front moves to higher positions when the angular velocity of the drum is growing (see Fig.~\ref{fig:averageFronts}). The increase in the front position $\theta_\text{f}$, follows the same proportionality to the square root of the angular velocity of the drum $\sqrt{\omega}$, see Fig.~\ref{fig:FluidFrontVsSpeed}, as long as the influence of particles is minimal. In a fluid-only scenario (black solid markers), the relationship $\theta_\text{f} \propto \sqrt{\omega}$ holds true for the whole range of simulated velocities. For simulations with particles, there is an initial linear trend, followed by an abrupt variation. This corresponds to the point when the particles get closer to the fluid front, which then becomes dominated by particle dynamics. For the limit of a very high particle content the linear trend is completely absent, and the behavior of the front can only be explained through the behavior of particles. 

At growing angular velocities of the drum the particle front moves in opposite direction to the fluid fronts, toward lower positions. This counter-intuitive effect can be explained by looking at the behavior of particles located close to the front.
The motion of these particles is governed by the relative magnitude of two time-scales. The first is the time of settling of a particle from its suspended state to the bottom layer, $t_\text{settling}$, and the second is the time it takes for the same particle to travel the distance between the grain bulk and the fluid front, $t_{\text{travel}}$. To get an estimate of $t_{\text{settling}}$, the solution of the equation of motion for an immersed spherical particle subjected to gravity and buoyancy~\cite{CourrechduPont2003,Cassar2005,Boyer2011} is considered:
\begin{equation}
\frac{\pi}{6}\rho_{\text{p}}d^3\frac{\text{d}u_r}{\text{d}t}=
\frac{\pi}{6} (\rho_{\text{p}}-\rho_{\text{f}}) g d^3-
\frac{\pi}{4\alpha} d \mu_0 u_r.
\label{eq::settling}
\end{equation}%

Here the flow is assumed to have a particle Reynolds number sufficiently low as to allow to use Stokes' law for the drag force, and the fluid plastic viscosity $\mu_0$ is used. The coefficient $\alpha$ takes into account the presence of surrounding particles. Its precise assessment is not easy in the drum condition. However, considering the particle packing to be close to the maximum, $\alpha$ can be estimated to be of order $10^{-2}$. The solution of Eq.~\ref{eq::settling} returns a settling velocity under stationary condition of
\begin{equation}
u_{r}=\frac{2}{3}\frac{\alpha (\rho_{\text{p}}-\rho_{\text{f}}) g d^2 }{\mu_0}.
\end{equation}

\begin{figure}[t]
 \centering
 \includegraphics[scale=1]{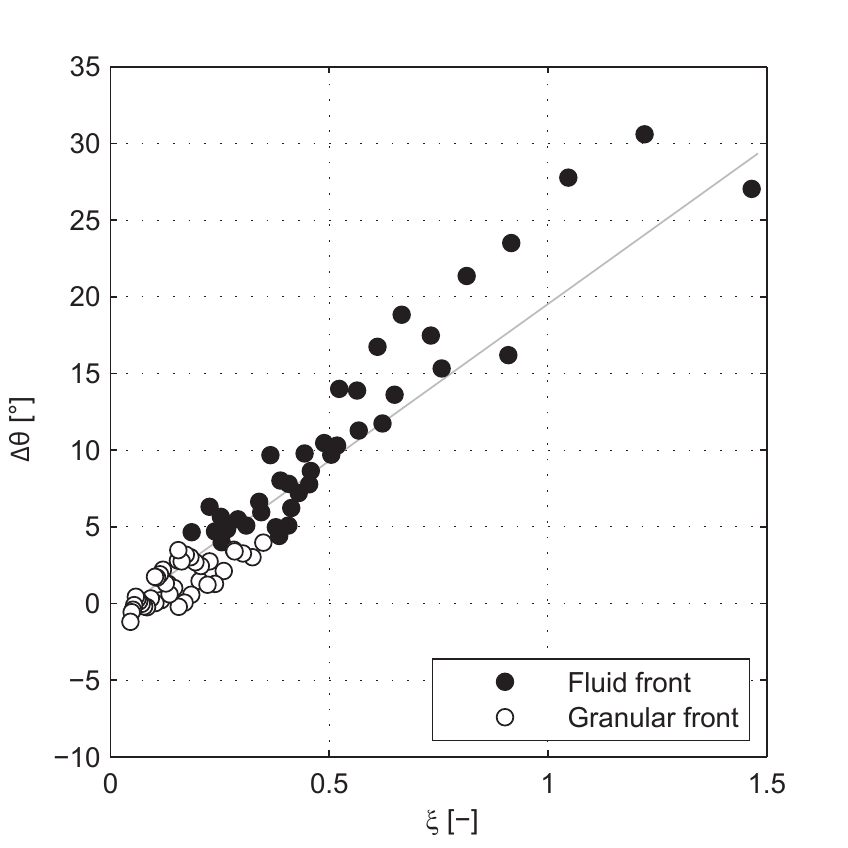}
 \caption{Angle between the particle bulk and the fluid front $\Delta \theta$ as a function of the time-scale ratio $\xi$, following a lineal trend (continuous line). Simulations where a granular front is observed ($\Delta \theta \leq 4^\circ$) are indicated by white markers.}
 \label{fig:DeltaThetaVsTimeRatio}
\end{figure}

A particle has to fall through a distance equal to the particle layer height, which can be assumed to be proportinal to  $\sqrt{V_\text{p} / W}$, with $V_{\text{p}}$ the total volume of particles and $W$ the drum width. The total settling time can be therefore estimated as
\begin{equation}
t_\text{settling} \simeq
\frac{h_{\text{p}}}{u_r} \simeq
\frac{3}{2} \frac{\mu_0 \sqrt{V_\text{p} / W} } {\alpha (\rho_{\text{p}}-\rho_{\text{f}}) g d^2 }.
\end{equation}
The same particle has a velocity in the $\theta$ direction of order $u_\theta=\omega R$ (see Fig.~\ref{fig:simulations}(b)), which is enforced by the boundary conditions and can therefore be considered to be independent from the viscosity of the fluid. This particle can potentially travel the distance between the particle bulk and the fluid front, $\Delta x$. This is related to the amount of free fluid, i.e. the fluid that is not trapped in the particle pores, $\Delta V=V_\text{f}-(1-\phi)V_\text{p}$, where $\phi$ is the particle volume fraction, which can be assumed to be close to its maximum value, $\phi_\text{max}=0.55$. Since $\Delta x \propto \sqrt{\Delta V / W}$, the travel time can be estimated as
\begin{equation}
t_\text{travel}=
\frac{\Delta x}{u_\theta} 
\simeq 
\frac{\sqrt{\Delta V / W}}{\omega R}.
\end{equation}%
The ratio between these two time-scales gives a dimensionless quantity able to express the tendency of a flow to develop a granular front. The definitions above yield
\begin{equation}
\xi=
\frac{t_\text{travel}}{t_\text{settling}}=
\frac{2}{3} \sqrt{\frac{\Delta V}{V_\text{p}}} \frac{\alpha (\rho_{\text{p}}-\rho_{\text{f}}) g d^2}{\mu_0 \omega R}.
\end{equation}
When $t_\text{settling}$ is bigger than $t_\text{travel}$, and therefore $\xi$ is small, particles will have time to reach the fluid front following the fluid streamlines, before the settling process leads them to the bottom of the flow and back into the particle bulk. Fig.~\ref{fig:DeltaThetaVsTimeRatio} shows the angle between particle front and fluid front $\Delta \theta = \theta_\text{f}-\theta_\text{p}$ plotted as a function of $\xi$. On this plot, the numerical data points aggregate around a line, which shows that the definition of $\xi$ well captures the  trend leading to the development of a granular front. Simulations with granular fronts ($\Delta \theta \leq 4^\circ$) are indicated by white markers, and correspond to a time-scale ratio lower than a threshold $\xi_\text{th}$ which lies in the range from $0.2$ to $0.4$ for this specific geometry.
%

\section{Conclusions}
\label{sec:conclusions}
We presented a series of experiments to study the behavior of a mixture of fluid and particles inside a rotating drum. Particles are heavier than the fluid and therefore have the natural tendency to settle. This is contrasted by enforcing a recirculation of the material through the rotation of the drum.  A sufficiently high angular velocities of the drum produces a front dominated by particles.  This reorganization pattern of fluid and particle phases is typical for suspensions, both natural (e.g. debris flow) and industrial (e.g. fresh concrete).
The same results are obtained in a numerical setting, using a hybrid method that combines LBM for the solution of the fluid phase with DEM for particle motion. Experiments and simulations show excellent agreement.

The segregation pattern is dependent on two phenomena. The first is the migration of the fluid front towards higher positions, which can be observed for growing angular velocities of the drum. This is explained by looking at the analytical solution for the free-surface flow of a Bingham fluid down an incline. At the same time, the particle front moves toward lower positions. The ratio between the time-scale of settling and the time-scale of particle recirculation defines a dimensionless quantity, $\xi$, that controls the formation of a granular front. The distance between particles and the fluid front is governed by $\xi$, which is proportional to the square root of the global particle content, and inversely proportional to the angular velocity of the drum. This result is confirmed by the measurements. When plotting  $\xi$ against the distance between particles and fluid front, numerical and experimental data collapse on a line. The time-scale ratio $\xi$ is proportional to the square of particle size and inversely proportional to the viscosity of the fluid, a tendency already observed in Ref.~\cite{Kaitna2011}. We were not able to test these further dependencies, and leave the proof to a later work. 
The validity of the $\xi$ scaling can be pivotal for the understanding of the the free-surface flow of particle suspensions. In particular, it can lead to better understanding of the condition under which a debris flow can develop, or a fresh concrete mixture will segregate during casting.

\begin{acknowledgments}
The research leading to these results has received funding from the People Programme (Marie Curie Actions) of the European Union's Seventh Framework Programme FP7 under the MUMOLADE ITN project (Multiscale Modelling of Landslides and Debris Flow) with REA grant agreement n$^{\circ}$ 289911, the ETH Zurich by ETHIIRA grant no.~ETH-03 10-3, as well as from the European Research Council Advanced Grant no.~319968-FlowCCS. The authors thank the \emph{Kamig \"{O}sterreichische Kaolin- und Montanindustrie} for providing the kaolin (Kamig E1) used in the experiments. Special thanks go to Dr. L. te Kamp and B. W\"{o}hrl for ongoing support withing the ITN.
\end{acknowledgments}

\bibliography{MendeleyAle,references}

\end{document}